# Resilient Decentralized Control of Inverter-interfaced Distributed Energy Sources in Low-voltage Distribution Grids

Alireza Nouri[1*], Alireza Soroudi[1], Andrew Keane[1]

[1] School of Electrical and Electronic Engineering, University College Dublin, Dublin, Ireland
[*]alireza.nouri@ucd.ie

**Abstract:** This paper shows that a relation can be found between the voltage at the terminals of an inverter-interfaced Renewable Energy Source (RES) and its optimal reactive power support. This relationship, known as Volt-Var Curve (VVC), enables the decentral operation of RES for Active Voltage Management (AVM). In this paper, the decentralized AVM technique is modified to consider the effects of the realistic operational constraints of RES. The AVM technique capitalizes on the reactive power support capabilities of inverters to achieve the desired objective in unbalanced active Low-Voltage Distribution Systems (LVDSs). However, as the results show, this AVM technique fails to satisfy the operator's objective when the network structure dynamically changes. By updating the VVCs according to the system configuration and components' availability, the objective functions will be significantly improved, and the AVM method remains resilient against the network changes. To keep the decentralized structure, the impedance identification capability of inverters is used to find the system configuration locally. Adaptive VVCs enable the decentralized control of inverters in an online setting. A real-life suburban residential LV-DS in Dublin, Ireland is used to showcasing the proposed method, and the effectiveness of proposed resilient active voltage management technique is demonstrated.

## 1. Introduction

Although the integration of Renewable Energy Sources (RESs) into the Low Voltage Distribution Systems (LVDS) have many positive effects if they are controlled efficiently, some negative impacts have been reported in the literature [1] for these and other Inverter-Interfaced Controllable Devices (IICDs). Power quality issues, e.g., over-/under-voltages and voltage unbalance were analysed in [2] for four low voltage feeders in Belgium. The authors concluded that although the integration of RESs, specifically photovoltaics (PV) in [2], may be beneficial to certain feeders, they may trigger power quality issues in others.

Reactive power scheduling and static voltage management in distribution systems is challenged by proliferation of Distributed Energy Resources (DERs), intermittent nature of RESs, increasing demand of the consumers and complexities associated with new types of loads. Though most DERs (including renewables) have the ability to control their output, growing penetration level of these sources increases the chance of conflict between their control actions and conventional control devices installed at different locations of distribution systems. Thus, these resources should be optimally schedule to hold an acceptable level of Power Quality (PQ) and also to reduce the operation cost. With no control scheme, the optimal operation will not be achieved and excessive operation of the regulating devices and other damaging phenomena are likely to happen [3]-[4].

The inverters connecting the IICDs have abilities other than converting DC power to AC. They are able to provide reactive power support [5]. Both European and American standards [6]-[7] allow DERs' reactive power support for voltage profile improvement. The control techniques can be classified to the centralized [8]-[9], decentralized and hybrid [10] approaches. Although the centralized and hybrid control schemes are able to effectively find the reactive power supports, they require significant investment on measuring, data collection and communication infrastructures. Alternatively different decentralized control schemes were proposed. In [11] the existing reactive power control schemes for IICDs were first analysed in balanced systems. A sensitivity analysis was then conducted to find the location-based reactive power strategies. A decentralized control was proposed in [12] for unbalanced feeders. Time series load flow calculations were performed using a range of predefined volt–Var, volt–watt, and power factor control settings.

A fully decentralized technique was proposed in [13] for optimal scheduling of reactive supports provided by IICDs in a LVDSs. The objective could be minimizing the voltage unbalance, energy loss or voltage deviation. This algorithm can be divided into two major steps, namely offline calculations and online application. In offline calculation step, first optimal voltages at Point of Common Coupling (PCC) of each IICD is found. If the optimal voltages are followed by the inverters, the operator's objective is best satisfied. Due to inverter limitations these voltages may be impossible to be accurately followed. Moreover, based on the standards, the single-phase voltage-sourced converters are not allowed to regulate the system voltage. However, they can be tasked with regulating a desirable amount of reactive power within their operational limits. Therefore, in the next step, for each scenarios, another optimization is conducted to find the optimal adjustment in the reactive power support provided by the IICDs. The decentralized control can be realized by finding a linearized relationship between the voltage measured at PCC and the optimal value of the reactive power support adjustment. This line is called Volt-Var curve (VVC). Section 2 discusses how to find the VVCs.

According to the case studies of this paper, the main factor that affects the VVCs is the system configuration. These means the network structure and also availability of the system controllable devices widely affect the VVCs. In other words, without tuning the VVCs, the voltage control objectives may never be realized. It is quite possible in operation of a LVDS that the system configuration is changed due to many reasons such as forced outages of the system components and operation strategies. These may drastically change the voltage intercepts of VVCs (target voltages at PCCs. In this paper, a simplified impedance identification technique is considered as an option to develop a more accurate and effective Active Voltage Management (AVM) scheme. The effects of the network configuration and also the availability of the system components on the AVM algorithm are analysed. The practical limitations of controllable devices are appropriately modelled and taken into account.

A decentralized AVM technique based on [13] is presented in Section 2. The modifications with respect to [13], the data required for the offline calculations and also the online implementation of the proposed AVM technique are discussed in this section. An AVM algorithm to consider the effects of the system configuration and components' availability is presented in Section 3. The results of the offline and online studies are presented in Section 4. The concluding remarks are presented in Section 5 based on these results validate the effectiveness of the proposed AVM method.

## 2. Summary of the Adopted AVM Technique

The basic foundations of the decentralized AVM algorithm based on [13] and the practical limitations that should be considered in the operation of IICDs are discussed in this section. The main idea is to find a VVC for each IICD using offline calculations. Then, in application mode, the IICDs are tasked with following the assigned VVCs to find the change in their reactive power support according to the measured voltages at regarding PCCs. A four-stage offline network analysis for obtaining VVCs is outlined below as a centralised solution. The decentralized implementation of the control technique is briefly discussed afterwards.

- **Stage I**: determines the optimal voltage across all scenarios that minimises the voltage unbalance of the feeder, or other objectives of interest, considering unlimited reactive power support for all RESs.
- **Stage II**: determines the change in the reactive power supports of inverter-interfaced devices which result in the closest possible voltage deviations from the optimal voltages found in stage I, in each scenario, forcing the reactive power support of the RES units to be within representatively realistic bounds.
- **Stage III**: the voltage levels are found without any changes in the reactive power injections of the controllable devices, i.e., base voltages.
- **Stage IV**: to conclude the offline-procedure, the resulting reactive power set-points (Stage II) are plotted against the resulting voltage set-points (Stage III) to determine the VVCs for each RES system.
- **Implementation of the control technique**: for each inverter, the voltage is measured at PCC and the value of the change in the reactive power injection is found using the regarding VVC. The value of the final reactive power injection of this inverter is found according to its capacity and other operational constraints.

The technique capitalises on the IICDs by engaging them in the provision of reactive power. Offline stages I-IV consist of a multi-scenario three-phase AC-OPF analysis.

The three-phase OPF tool employs a mathematical technique (optimization solver) to navigate the solution space outlined by the problem formulation. A converged solution ought to satisfy the equations governing power flow. In stage I, this optimization is provided in (1)-(8). The objective can be loss minimization, voltage profile improvement, voltage unbalance improvement or a combination of these objectives. In (1), $N_s$ is the number of scenarios indexed by $s$. $V_i^{\text{opt}}$ is the optimal voltage that should be followed by inverter $i$, to achieve objective (1). The voltage between phase $\kappa$ and neutral ($n$) at bus $b$ in scenario $s$, should be within the predefined limits (2). The current of line $l$ on phase $\rho$ in scenario $s$ should be lower than the maximum allowable limit (3). Voltage-dependent nature of the loads is modelled using ZIP decomposition method which combines the effects of constant power, constant current and constant impedance load components. $\alpha$, $\beta$, $\gamma$ are the coefficients of ZIP model. Each scenario vector includes the characteristics of the load model ($\alpha$, $\beta$, $\gamma$, $Pd_0$ and $Qd_0$) and active power production of IICDs, i.e., $Pg_i$ for inverter $i$. Equation (5) gives the total active and reactive power consumption at bus $b$, where $\text{IICD}_b$ is the set of IICDs connected to bus $b$. The apparent power relationship is given in (6) for bus $b$. Equations (7) and (8) enforce KCL and KVL, respectively, where buses $m$ and $n$ are the sending and receiving ends of line $l$.

$$\text{Min } F_I = \sum_{s=1}^{N_s} f_s(V_i^{\text{opt}}) \tag{1}$$

$$\underline{V_b} \leq |V_{b,s}^{\kappa n}| \leq \overline{V_b} \qquad \forall b, s, \kappa \in \{a,b,c\} \tag{2}$$

$$\underline{I_l} \leq |I_{l,s}^{\rho}| \leq \overline{I_l} \qquad \forall s, \forall \rho \in \{a,b,c,n\} \tag{3}$$

$$Pd_{b,s}^{\varphi\varphi'} = Pd_{b,s,0}^{\varphi\varphi'} \left( \alpha_{b,s}^{\varphi\varphi'} \frac{|V_b^{\varphi\varphi'}|^2}{|V_{b,0}^{\varphi\varphi'}|^2} + \beta_{b,s}^{\varphi\varphi'} \frac{|V_b^{\varphi\varphi'}|}{|V_{b,0}^{\varphi\varphi'}|} + \gamma_{b,s}^{\varphi\varphi'} \right)$$

$$Qd_{b,s}^{\varphi\varphi'} = Qd_{b,s,0}^{\varphi\varphi'} \left( \alpha_{b,s}^{\varphi\varphi'} \frac{|V_b^{\varphi\varphi'}|^2}{|V_{b,0}^{\varphi\varphi'}|^2} + \beta_{b,s}^{\varphi\varphi'} \frac{|V_b^{\varphi\varphi'}|}{|V_{b,0}^{\varphi\varphi'}|} + \gamma_{b,s}^{\varphi\varphi'} \right) \tag{4}$$

$$P_{b,s}^{\kappa} = Pd_{b,s}^{\kappa} - \sum_{i \in \text{IICD}_b} Pg_{i,s}^{\kappa}$$

$$Q_{b,s}^{\kappa} = Qd_{b,s}^{\kappa} - \sum_{i \in \text{IICD}_b} Qg_{i,s}^{\kappa} \tag{5}$$

$$V_{b,s}^{\kappa n} \left( I_{b,s}^{\kappa} \right)^* = P_{b,s}^{\kappa} + jQ_{b,s}^{\kappa} \tag{6}$$

$$I_{b,s}^{\kappa} = \sum_{l \in \text{Line}_b} I_{l,s}^{\kappa}$$

$$I_{l,s}^{n} = I_{l,s}^{a} + I_{l,s}^{b} + I_{l,s}^{c} \tag{7}$$

$$V_{m,s}^{\kappa n} - V_{n,s}^{\kappa n} = R_l I_{l,s}^{k} + j \sum_{\rho \in \{a,b,c,n\}} X_l^{k\rho} I_{l,s}^{\rho} \tag{8}$$

The solution of this optimization gives the optimal voltages that should be followed by the inverter controllers ($V_i^{opt}$). However, According to the standards and network codes, the single-phase voltage-sourced converters are usually neither equipped with voltage controllers nor permitted to regulate the system voltage [14]. They can regulate their reactive power supports. Therefore, instead of following a voltage set point, each inverter is tasked with injecting (or absorbing) a value of reactive power which pushes the PCC voltages towards the optimal voltages. In stage II, another optimization is conducted separately for each scenario. Using this optimization, the change in the reactive power support provided by all inverters are found in a way that minimize the average deviation of the PCC voltages from the optimal voltages found for each scenario. $N_s$ separate optimizations are conducted. The objective function of these optimizations is provided in (9). The operational constraints of the inverters are also taken into account (10)-(12).

$$Min \; F_{II}(\Delta Qg_{i,s}) = \sum_{i=1}^{N_{IN}} (V_{i,s} - V_i^{opt})  \quad (9)$$

**S.t.** (2)-(8) and:

$$-\sqrt{\overline{S}_i^2 - Pg_{i,s}^2} \leq Qg_{i,s} + \Delta Qg_{i,s} \leq \sqrt{\overline{S}_i^2 - Pg_{i,s}^2} \quad (10)$$

$$-\tan(\overline{\alpha_i})Pg_{i,s} \leq Qg_{i,s} + \Delta Qg_{i,s} \leq \tan(\overline{\alpha_i})Pg_{i,s} \quad (11)$$

$$-\overline{Q_i} \leq Qg_{i,s} + \Delta Qg_{i,s} \leq \overline{Q_i} \quad (12)$$

For stage III, a separate power flow analysis is conducted for each scenario to find the PCC voltages without any change in the reactive power supports ($\Delta Qg_{i,s}=0$ for all inverters). The output of stage III for each inverter $i$, is a vector of base voltages for all scenarios ($V_{i,s,base}$). In stage II, the optimal changes in the reactive power support of each inverter $i$ in each scenario ($\Delta Qg_{i,s}$) were found. $\Delta Qg_i$ and $V_{i,base}$, for all scenarios are the inputs of stage IV for inverter $i$. This two vectors are first plotted against each other. Then, a linear relationship is found for $V_{i,base}$ and $\Delta Qg_i$ using linear regression technique. Fig. 1 shows a sample of such a linear relationship. This line is the VVC which can be used in decentralized control of inverters. Based on [13], the voltage intercept of VVC should be close enough to the optimal voltage found in stage I for this inverter.

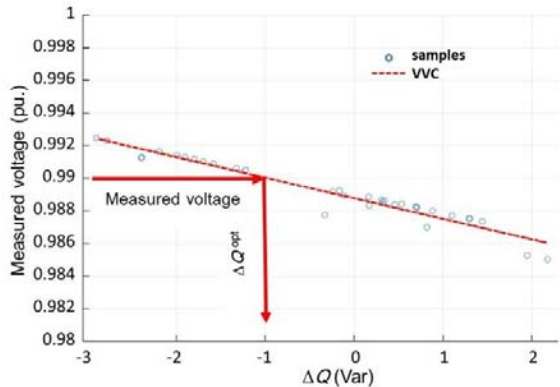

***Fig. 1.*** *Sample VVC for decentralized voltage control.*

Before the final value of reactive power supports found using VVCs be applicable, the operational constraints should be taken into account in stage V. In most of the practical applications, a simple power factor limit, e.g., power factor>0.95 lag, is still being used as the industry common practice for reactive power control. In this paper, more accurate operational constraints (which are mostly proposed for the safe operation of RESs as well as achieving higher levels of power quality) are considered. It is discussed in the case studies, how these constraints affect the voltage controllability comparing to the common industry practice. Fig. 2 presents the operational constraints of a typical inverter.

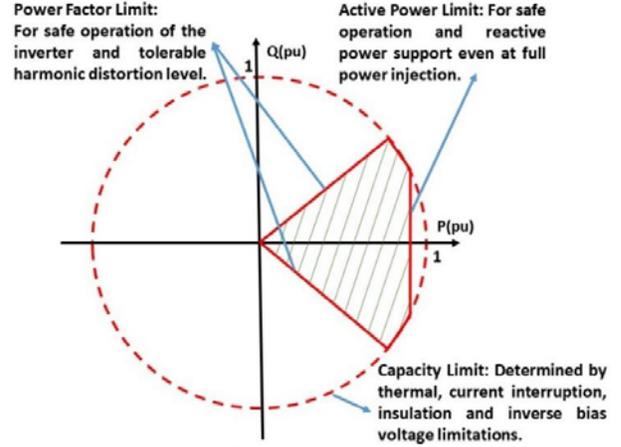

***Fig. 2.*** *Typical Capability (P-Q) Curves for LV/MV Inverters.*

This paper deals with AVM which can be interpreted as static voltage control in active LVDSs. Even though the method can be applied to control the system voltages with as high resolution as the operator has in mind, in the short time scales i.e., milliseconds, some other effects may need to be taken into account. Dynamic voltage stability issues may arise, for which the dynamic voltage control techniques should be applied. One of these techniques is the one recently presented in [15] based on the inverters' impedance measurement capability. Dynamic control techniques cannot be used for static voltage control. The ZIP model is used for system loads in the proposed AVM algorithm. The ZIP coefficients of the loads depend on many factors. For instance, the ZIP coefficients of motor loads depend on motor type, size, load and speed. However, this has been already considered in the proposed method, since the ZIP coefficients are considered among the uncertain parameters in the scenarios and the proposed AVM algorithm is resilient against the variations of these parameters.

### 3. Proposed Resilient AVM Technique

A decentralized AVM algorithm was discussed in Section 2. In the case studies, this algorithm is first tested on a real-life LVDS to demonstrate its performance for effective voltage control in unbalanced LVDS with high penetration of IICDs. This high penetration leads to both new challenges (such as voltage deviation and reverse power flow), and also new opportunities for control of these systems.

According to results, one of the main factors that affects the Volt-Var curves in a three-phase unbalanced

LVDS is the system configuration and availability of the system controllable devices. It is quite possible that in operation of a LVDS, the system configuration is changed due to faults, scheduled maintenance, forced outages of the components and operation strategies. These change the parameters of VVCs, especially the voltage intercepts of VVCs. As will be discussed in the case studies, this intercept is approximately equal to the target voltage at PCCs which is the voltages that the inverter should follow.

AVM algorithm should be modified to include the effects of variability of the grid configuration and also availability of the system controllable devices. In this paper, it will be shown that without an adaptive AVM framework the voltage control objectives cannot be realized. However, based on just voltage measurement and without a strong communication system to exchange the data between the decentralized control agents, it is not impossible to keep the parameters of VVCs up to date. To keep the decentralized structure of the AVM technique, a methodology is presented here to tweak the parameters of AVM algorithm to best comply with the system topology and components' availability based on the available impedance measurement capability (subsection 3.1) of the current inverters.

### 3.1. Impedance Identification Technique

The ability of the system inverters to measure the grid impedance is important in order to evaluate the stability of the power electronic interface. Offline impedance identification techniques are not viable in practice due to the ever-changing nature of grid impedance over time. The online methods of impedance measurements are being used in order to monitor system stability in a decentralized manner in real time and synthesize corrective actions in case they are necessary [15].

Impedance measurement techniques based on sine sweeps are not applicable for online impedance measurements. The other important feature is that the impedance measurement process should be non-invasive. There are many invasive techniques reported in the literature which require additional power amplifiers and switching passive loads [16]. Such techniques are usually difficult to implement and very expensive. The impedance measurement technique which is applied here requires no additional power level hardware. This technique was well discussed in [15]. Non-invasive Wideband System Identification (WSI) techniques are able to inject the signals which excite the system at all frequencies, allowing impedance measurement for a wide range of frequencies within a very short period [15]. Such impedance identification techniques were initially proposed for DC applications [17]. With some modifications, these techniques have been extended to the AC systems.

In the technique proposed in [15], which is also used here, a set of signals that have characteristics of white noise for the frequency range of interest is imposed on the control signal or more accurately the reference signal in the control unit of the inverter [17]. Synchronous voltage and current measurements are recorded by a dedicated metering unit within the converter controller. A non-parametric impedance function is then formed. A linear impedance identification technique is then applied to get the parametric impedance model. In [15], the Pseudo Random Binary Sequence (PRBS) white noise was applied. Such signal can easily be generated using the shift-register concept. The measured impedance is used here to find the steady state configuration and components' availability. The above-mentioned technique injects perturbation signals when the system is in a steady state condition. Furthermore, the steady state impedance identification techniques are preferred since they only inject small signal perturbation whereas transient methods perturb the grid way more aggressively.

In this technique, there are three different options for injecting the PRBS noise. The simulated white noise signal can be injected in the current reference or in the controllers' output. The other option is to inject the noise in current reference and the controllers' output simultaneously. The converter controller first decomposes the control signals for $d$ and $q$ axes. The PRBS noise signal can also be decomposed and subsequently superimposed on the control signal on these axes. The tuned parameters of the AVM algorithm are going to be applied for a relatively long time horizon, i.e., as long as the measured impedance change is limited. The PRBS signal is firstly injected along $d$-axis and the d-axis impedances are measured. After measuring the measurement impedance, the PRBS noise signal is injected along the controller $q$-axis and the associated impedances are measured. Both $d$ and $q$ axes control signals can simultaneously be perturbed and the impedances can be measure after getting rid of cross-correlation using the state-of-the-art methods [15].

### 3.2. Resilient AVM Algorithm

The futuristic LVDSs include various RES technologies. The Thevenin equivalent of the network seen by each individual RES is different and depends on the network characteristics and also the behaviour of other RESs in the network as shown in Fig. 3. The Thevenin impedance seen by each RES depends on several factors such as those as described in Fig. 4. Among these factors, the impact of network parameters, as well as the demand characteristics on VVC, are already captured in operating scenarios (Section 2).

The VVC of each RES is obtained and used for AVM, as described in Section 2. The idea of applying VVCs is based on the fact that the equivalent Thevenin model of the network seen from PCC of each RES does not widely change. Many factors might cause changing this Thevenin model. This will reduce the effectiveness of the AVM algorithm. Fig. 5 shows influence of RES/inverter failure on AVM performance is.

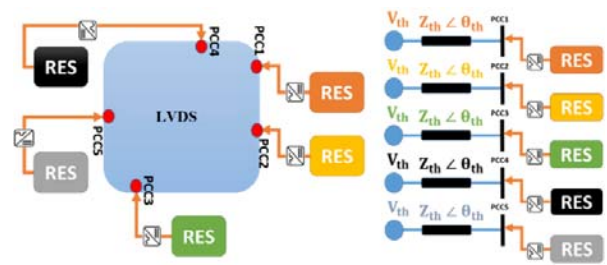

***Fig.3.*** *Schematic of network seen by each individual RES.*

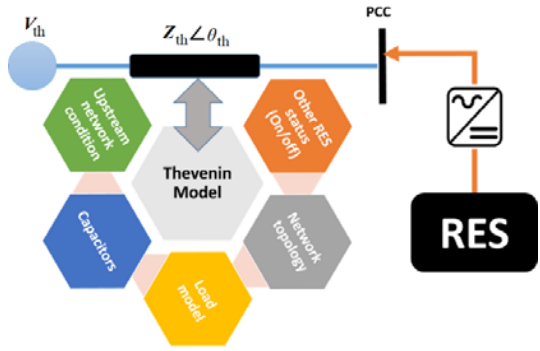

***Fig. 4.*** *Factors influencing the equivalent Thevenin model seen by each RES at its PCC.*

Fig. 5-a) shows the futuristic distribution network with several inverter based RES connected to the grid. These technologies can be of PV, wind, storage or any inverter-based energy resources. Fig. 5-b) shows the same futuristic LVDS. For each RES, a VVC is tuned. Fig. 5-c) Shows the case when there is a contingency on one of the RES units. In this case, the pre-tuned VVCs may not able to provide the required performance. This is due to the fact that the Thevenin model seen by each RES is changed and the VVC is no longer valid. In Fig. 5-d), if there is any contingency on any of the RES, initially, the contingency is identified for each RES using a local impedance identification technique (subsection 3.1), the VVCs for the rest of them will then be updated to capture the changes in the network topology. This resilient AVM technique remains robust against the contingencies. The framework for VVC extraction in resilient AVM method is shown in Fig. 6. The following assumptions are made to obtain the resilient set of VVCs. The only contingency considered in this framework is the failure of the inverter based RES units. Only single contingencies are captured. At contingency row $i$, it is assumed that the inverter at RES $i$ is failed and is not able to inject power to the grid.

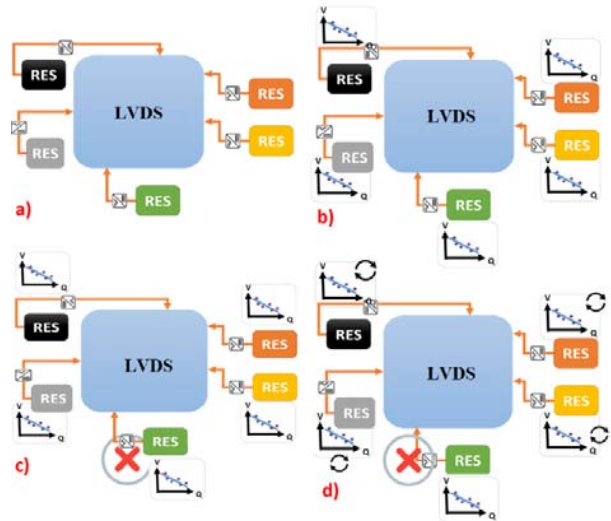

***Fig. 5.*** *Influence of the RES failure on AVM performance.*

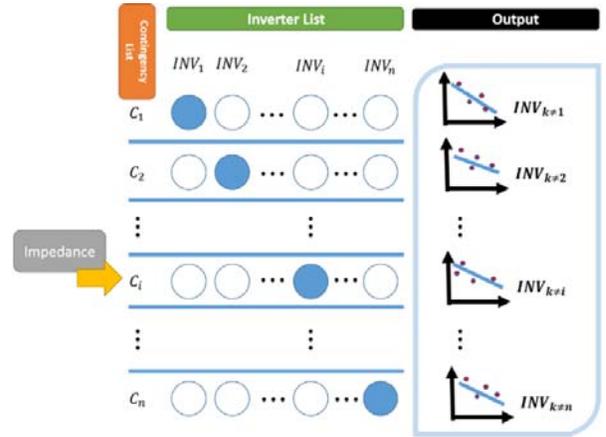

***Fig. 6.*** *Schematic resilient VVC determination for AVM.*

Once the VVCs shown in Fig. 6 are found, the utilization of the Resilient AVM (RAVM) algorithm is straight forward. The impedance seen from the PCC of each RES indicates which row of Fig. 6 should be used for selecting the appropriate set of VVCs.

## 4. Case Studies

The sample LVDS which is used to showcase the AVM technique is a radial LV feeder with 85 nodes situated in Ireland [18]-[19]. A 10/0.4 kV step-down transformer with fixed tap position is used to connect this 11-node 74-customer underground network (Fig. 7) to the upstream system. To ensure realistic voltage perturbations are observed at the head of the feeder in the multi-scenario case, a separate feeder connection off the transformer is considered as presented in Fig. 7, which supplies further 85 customers. In order to be able to plot the three-phase voltages, the buses are renumbered in Fig. 7. The RESs available in this case study are inverter-based 2 kW V2G systems. Fig. 7 also presents the phase that these V2G inverters are connected to as well as their locations and three-phase supply nodes. A high resolution (minute by minute) database on the active and reactive power demands at each bus is available. In order to reduce the computational burden of the proposed AVM algorithm, simultaneous backward scenario reduction algorithm is applied to reduce these huge number of samples to a few number of scenarios [20]. Fig. 8 shows the active power consumption during a day by a one-minute resolution. Fig. 9 presents the active power consumption for each scenario. These scenarios model the actual load variations with an acceptable accuracy. This show the effectiveness of the adopted scenario reduction algorithm. Other input parameters, e.g., reactive power demands and ZIP model parameters [21], are also included in scenarios (Section 2).

Batteries of Electric Vehicles (EVs) have a considerable potential not only to provide energy for the locomotion of EVs, but also to dynamically interact with the LVDSs. Thereby, through the energy stored in the batteries, these vehicles can be used to regulate the active and the reactive powers. It should be noted that low voltage V2G systems typically consume active power. The owners of the electrical vehicles may raise an argument about discharging their vehicles' batteries in the course of time that they left their vehicles to be charged. Here, it has been assumed that the charging station only consumes active power. However, reactive power can be easily exchanged between the V2G system and the grid. To elaborate V2G inverters can work in all four quadrants of P-Q plane (see Fig. 2). For the aforementioned reasons, it is assumed that the V2G systems only absorb active power. The V2G system can increase the reactive power injected to the grid by increasing the voltage at the AC terminal of the inverter with respect to the voltage at PCC and can also absorb more reactive power by decreasing the terminal voltage with respect to the voltage at PCC. It should be noted that the proposed algorithm for decentralized AVM can also work with no change under bidirectional active power exchange assumption.

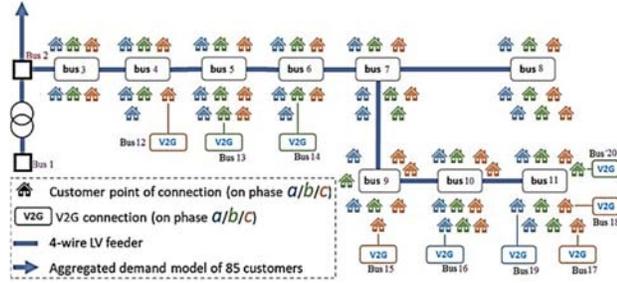
*Fig. 7.* Topology of the sample LVDS used for the case studies.

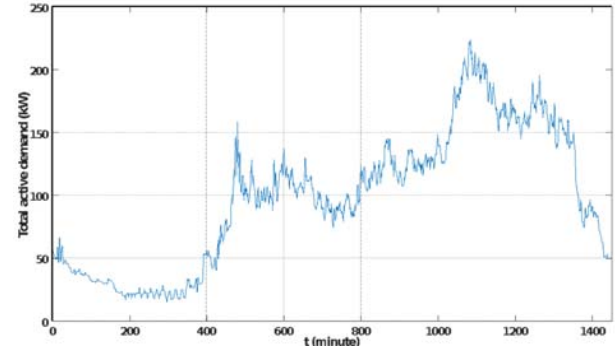
*Fig. 8.* Minute-by-minute total active power consumption.

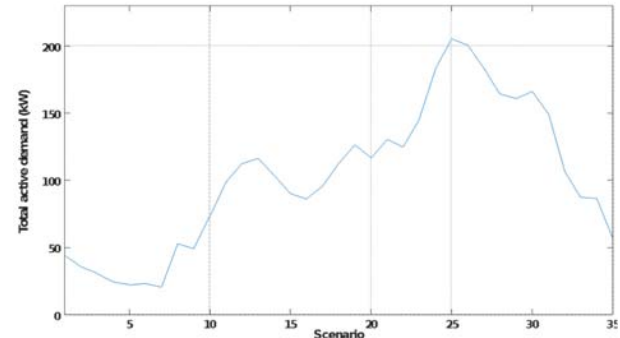
*Fig. 9.* Total active power consumption in each scenario.

### 4.1. Modified AVM Algorithm Based on VVCs

First, the results are presented assuming that there is no limitation on the reactive power provided by these inverters except for the inverter maximum capacity ($S^{max}$). The offline modelling phase is repeated for voltage unbalance minimisation as the first objective and loss minimisation as the second objective. The VVC of each V2G system is found using the method discussed in Section 2.

All the slopes of the VVCs should be negative. This assures a positive reactive power injection when the measured voltages at PCCs are below the regarding target voltages ($V^{opt}$) and a negative reactive power injection (positive absorption) when the measured voltages are higher than the target voltages. This validates the successful application of the proposed AVM algorithm. The target voltage (voltage intercept $c$), should closely match the optimal voltage ($V^{opt}$), i.e., the voltage set-points extracted for voltage control mode. This indicates the effectiveness of the proposed method for AVM based on VVCs. Here the relative error is defined as the relative difference between the intercept of each VVC and the optimal voltage ($|c-V^{opt}|/V^{opt}$). In none of the cases, the relative error exceeds 1%.

For the first objective, the sub-plots of Fig. 10 are the reactive power set-points from the second stage in the AVM algorithm and the voltage set-points from the third stage plotted against each other and also the VVC for each respective V2G system. For each subplot, the VVC equation for reactive power control is passed to the online implementation phase. For the second objective, the relative errors are less than 0.1%. The slopes of all VVCs are also negative. Fig. 11 shows the PCC voltages plotted against the optimal values of the reactive power injection for each V2G system in 35 scenarios (after scenario reduction) and also the VVCs. The optimal reactive power injections and voltages are found according to the method presented in Section 2.

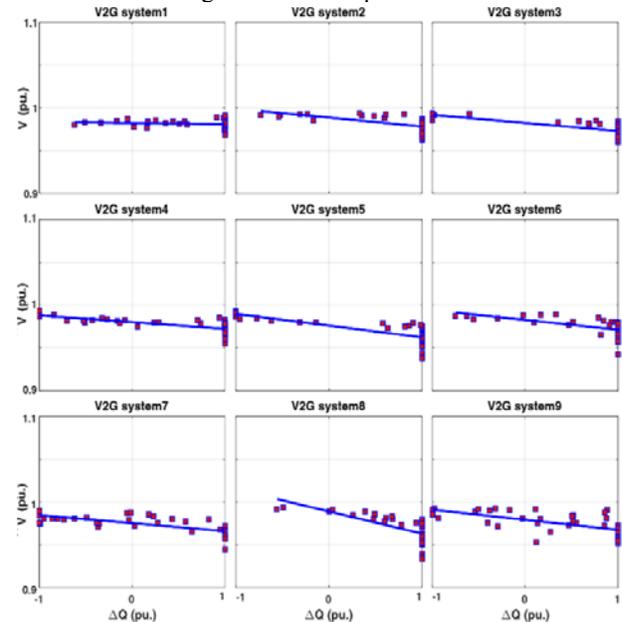
*Fig. 10.* Resulting VVCs for V2G systems, showing intercepts and slopes, objective: minimisation of voltage unbalance.

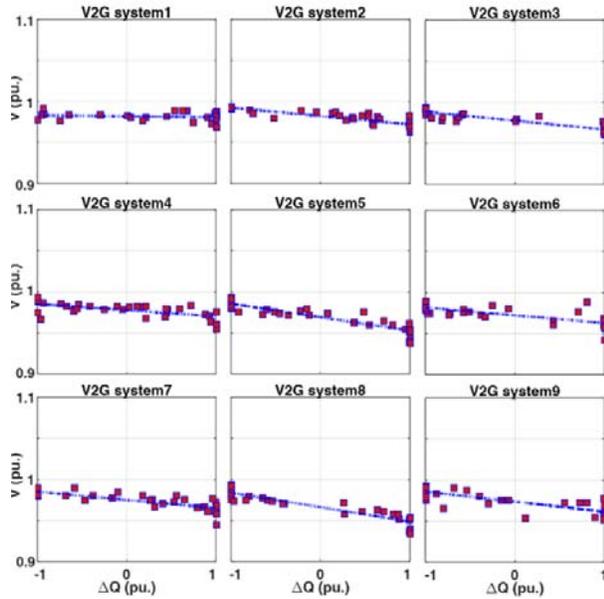

*Fig. 11. Resulting VVCs for V2G systems, showing intercepts and slopes, objective: minimisation of total loss.*

As a measure of goodness-of-fit to show how close the data are to the fitted regression line, R-squared test is done for all the VVCs extracted in the case studies. R-squared measure is the percentage of the response variable variation that is explained by the linear model. For 73% of the VVCs, this measure is higher than 70%. Even though the interpretation of R-squared measure is hard, such level of this measure assures the validity of the VVCs. If after conducting the linear regression, the value of this measure was less than expected, the boundary scenarios and outliers in the scenario set were removed and the linear regression was conducted again. In the case that the linear regression does not pass the R-squared test, more complicated curves can be fitted to the data or other control schemes should be followed.

Comparing the VVCs obtained for loss minimization to those obtained for minimization of the voltage unbalance, one can say the intercepts of the VVCs obtained for loss minimization in this system are usually lower, since due to the load-to-voltage dependence, the higher voltages lead to the higher active and reactive power demand. Higher active and reactive powers lead to higher line currents, which in turn cause higher power loss. No meaningful relation was observed for the slopes of the VVCs for these two objectives.

To validate the ability of the adopted 3-phase power flow algorithm with many controllable devices connected to the LVDS under study, and to assess the voltage controllability of the inverters, the three-phase voltages at all system buses are presented in Fig. 12. The voltage set-points of each inverter is fixed on the target values found under voltage unbalance minimisation as the main objective. Considering the voltages at buses 12-20 (PCC of V2G systems), it can be seen that the voltages of the phases on which the V2G systems have been installed match the target values presented Fig. 10. The same study is conducted under loss minimization as the main objective. The three-phase voltages are depicted in Fig. 13. As discussed earlier, the optimal voltages are usually lower with this objective. The adopted unbalanced three-phase power flow algorithm is effectively converged to stable solutions.

With the optimal voltages obtained for voltage unbalance minimisation, the voltages at system load points (bus 1-11) are closer to each other. This leads to a lower voltage unbalance and shows the effectiveness of the optimisation algorithm used to find the optimal voltages. To analyse the effectiveness of the AVM technique, the minute by minute active and reactive power demands at all load points and also the other required data are collected for one week and for each minute a three-phase unbalance power flow has been conducted to find the three-phase voltages. In this week-long time-series power flow, the V2G systems on this LV feeder are tasked with following their assigned VVCs.

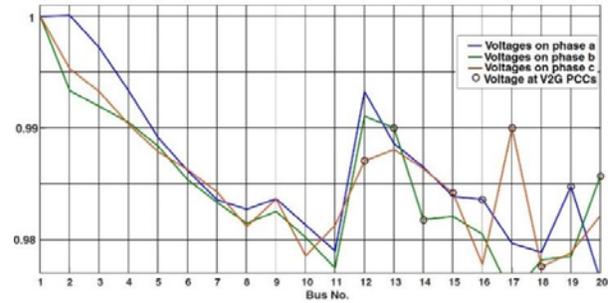

*Fig. 12. Bus voltages, (Minimisation of Voltage Unbalance).*

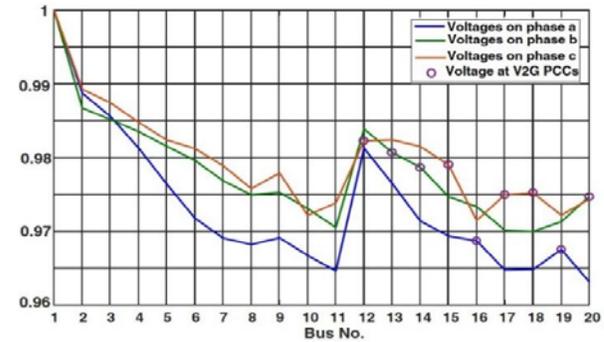

*Fig. 13. Bus voltages, (Minimisation of total loss).*

Table 1 summarizes the average performance metrics found using the minute-by-minute one-week-long time-series power flow simulations utilising the VVCs found in the offline-analysis for voltage unbalance and loss minimization. The results are also presented for the fixed power factor scenarios. With minimisation of the voltage unbalance as the objective, the algorithm reduces the voltage unbalance by about 26% and 37% comparing to the voltage unbalance found with loss minimization as the main objective and the best voltage unbalance gained with the fixed power factor assumption. On the other hand, under loss minimization objective, the weekly energy loss is reduced by 11% and 5% compared to the energy loss with minimization of the voltage unbalance as the main objective and the best voltage unbalance found under the fixed power factor assumption.

The remaining part of this subsection discussed the effects of the other limitations on the operation of the inverter-based controllable devices on the performance of the proposed active voltage management algorithm based on VVCs. Table 1 also presents the results obtained in the implementation of the VVCs for AVM considering two different set of constraints on the operation of the inverter of the V2G systems to show the effects of the constraints.

In one study, it is assumed that the maximum reactive power support that can be provided by each inverter has a fixed component, e.g., 30% of the maximum capacity which can be provided even when there is no active power injection or absorption. There is also an active power-dependent component for the maximum reactive power support that is proportional to the value of the active power injection/absorption of the regarding inverter. We model this component by applying a limit on the value of the power factor of this inverter, e.g. $PF^{max}=0.82$. In Table 1, this study is referred to as the one with accurate operational constraints.

In another study, a maximum power factor is set for the inverter operation. In this study the inverters can only absorb the reactive power signifying an always lagging power factor. This is the common engineering practice in most of Distribution System Operator (DSOs) across Europe. Here this maximum lagging power factor is set to 0.92. It should be noted that in this study, the power factor is not assumed to be fixed. The value of the reactive power injection of each inverter is still found using the regarding VVC, but the maximum reactive power injection is restricted by this maximum power factor assumption.

Now it can be discussed how these constraints affect the performance of the proposed AVM algorithm. According to Table 1, with the accurate constraint modelling, the results of applying the proposed AVM algorithm for optimising the reactive power dispatch to minimise the voltage unbalance are even better than the case with capacity constraint as the only constraint on the operation of the system inverters. It may seem a little bit surprising, but can be justified as follows.

1) The performance metrics will be improved (compared to the fixed power factor and uncontrolled dispatch). However, the results are not globally optimal. Actually, the globally optimal results cannot be achieved using any decentralized control system. The results cannot be treated as the results of an optimization framework that is applied to find the global optimal settings.
2) After applying the VVCs to find the optimal reactive power injection of all inverters, the set of voltage measurements will be updated. It is quite possible that according to these new voltage levels, further corrections are required. This indicates the need of applying a supervised closed loop voltage control based on the VVCs that is out of the scope of this paper.

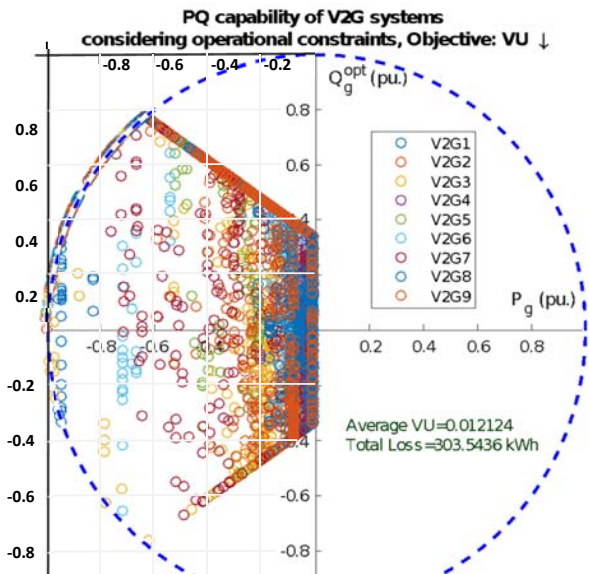

*Fig. 14.* Active and reactive power injected by each inverter during the week of study (with actual operational constraints).

**Table 1** Energy loss and voltage unbalance metrics for different cases

| Objective | Total Loss (kWh) | Average V. Unbalance (%) |
|---|---|---|
| Min. V. Unbalance ($S^{max}$ constraints) | 313.44 | 1.26 |
| Min. V. Unbalance (accurate constraints) | 303.54 | 1.21 |
| Min. V. Unbalance Lagging PF>0.92 | 297.65 | 1.52 |
| Min. Power Loss [kW] | 283.95 | 1.58 |
| Min. Power Loss (accurate constraints) | 282.98 | 1.52 |
| Min. Power Loss Lagging PF>0.92 | 290.48 | 1.59 |
| Fixed 0.95 Lag Power factor | 297.15 | 1.74 |
| Fixed Unity Power factor | 298.95 | 1.72 |
| Fixed 0.95 Lead Power factor | 301.11 | 1.72 |

In three separate studies, three different fixed power factor operation strategies are also assumed for the V2G systems, i.e., 0.95 inductive, 1 and 0.95 capacitive. A fixed power factor is typical for an inverter based controllable device connected to a LVDS to reduce the voltage-rise effect caused by the excessive active power injections. The results of applying the VVCs to find the set-points are compared to the results of this fixed power factor strategies.

With total energy loss as the objective, the value of weekly energy loss is lower considering the actual constraints on the operation of the system inverters comparing to the energy loss obtained for the optimised reactive power dispatch considering the capacity limit as the only constraint.

For the study with maximum lagging power factor of 0.92, the proposed AVM algorithm leads to a total loss and voltages unbalance of 2.5% and 12.5% lower than the best energy loss and voltage unbalance attained with the fixed power factor assumption with loss minimization and minimization of the voltage unbalance as the main objective, respectively. This indicates the effectiveness of the proposed AVM algorithm even with such restricted feasible region.

To show how the actual constraints restrict the reactive power support provided by each inverter in this study, Fig. 14 shows the values of the active and reactive power injections of all 9 V2G inverters in this weeklong study with minimization of the voltage unbalance as the objective.

### 4.2. Resilient AVM Algorithm

To obtain the VVCs in case of the outage of each inverter a centralised offline simulation is conducted. For each possible connection state, a set of VVCs are extracted. After extracting the VVCs in each connection state, the results of applying the proposed adaptive AVM technique are compared with those obtained by applying the fixed VVCs.

Here, all V2G systems are disconnected one by one and a set of VVCs are extracted for the system inverters. It is not possible to show the VVCs for all connection modes. Instead, Fig. 15 shows the optimal voltages found for the remaining inverters in the case of the N-1 outages. As shown in Fig.15, the optimal voltage levels at PCCs of the system inverters depend on the availability of the other system inverters. This indicates the necessity of developing an adaptive decentralised control scheme.

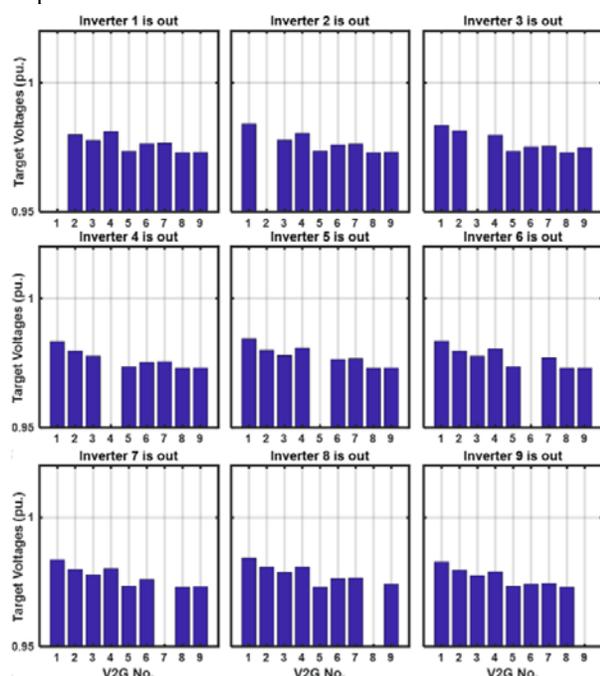

***Fig. 15.*** *Optimal voltages of the other inverters after outage of each V2G system (minimization of voltage unbalance).*

The effectiveness of the adaptive voltage control framework should be validated. To this end, two new studies are conducted. For both studies the proposed AVM algorithm is implemented in a one week period to keep the reactive power supports up-to-date (see subsection 4.1). During each day of the study week a certain inverter is assumed to be unavailable. V2G systems 8 and 9 are always available.

First study: In the week-long time-series power flow, the V2G systems are tasked with following their assigned VVCs found in subsection 4.1. In other words, the VVCs are not updated according to the availability of the V2G systems.

Second study: In a week-long horizon, the V2G systems are tasked with following their assigned VVCs. It means during each day, the set of VVCs are updated according to the outage of other V2G systems.

**Table 2** Energy loss and voltage unbalance (first study)

| Objective | Total Loss (kWh) | Average V. Unbalance (%) |
|---|---|---|
| Min. V. Unbalance ($S^{max}$ constraints) | 301.02 | 1.49 |
| Min. V. Unbalance (accurate constraints) | 302.21 | 1.39 |
| Min. V. Unbalance Lagging PF>0.92 | 296.79 | 1.51 |
| Min. Power Loss [kW] | 288.92 | 1.62 |
| Min. Power Loss (accurate constraints) | 287.80 | 1.55 |
| Min. Power Loss Lagging PF>0.92 | 290.86 | 1.62 |
| Fixed 0.95 Lag Power factor | 297.15 | 1.74 |
| Fixed Unity Power factor | 298.95 | 1.72 |
| Fixed 0.95 Lead Power factor | 301.11 | 1.72 |

**Table 3** Energy loss and voltage unbalance (2nd study)

| Objective | Total Loss (kWh) | Average V. Unbalance (%) |
|---|---|---|
| Min. V. Unbalance ($S^{max}$ constraints) | 308.42 | 1.26 |
| Min. V. Unbalance (accurate constraints) | 305.23 | 1.22 |
| Min. V. Unbalance Lagging PF>0.92 | 298.75 | 1.50 |
| Min. Power Loss [kW] | 284.83 | 1.65 |
| Min. Power Loss (accurate constraints) | 283.09 | 1.58 |
| Min. Power Loss Lagging PF>0.92 | 289.48 | 1.67 |
| Fixed 0.95 Lag Power factor | 297.15 | 1.74 |
| Fixed Unity Power factor | 298.95 | 1.72 |
| Fixed 0.95 Lead Power factor | 301.11 | 1.72 |

Tables 2 and 3 show the system energy loss and voltage unbalance for the first and second studies, respectively. Each study has been repeated for the objectives of minimisation of the voltage unbalance and loss minimisation with different assumptions for the limitations that should be considered for the reactive power support capability of the system inverters. The details of such assumptions are provided in subsection 4.1.

As can be seen in Table 2 and 3, with the objective of minimisation of the voltage unbalance, the value of the optimal voltage unbalance index is reduced at least by 14% in the second study compared to the first study. The value of the total energy loss is reduced at least by 1.66% in the second study comparing to the first study when the loss minimisation is chosen as the system level objective function.

## 5. Discussion and Concluding Remarks

According to the results, the VVCs enable the decentral operation of inverter-based control devices for voltage control in an online setting satisfying the performance criteria. The optimal target voltages of inverters should be identified in a wide range of diverse conditions including the entire range of demand-to-voltage sensitivities. Even with these diverse conditions, the optimal voltages can be found to effectively satisfy the DSO's objective. Another point of note is revealed when considering the small relative errors between the optimal voltages and the intercept of the VVCs. In fact, a small relative error indicates that the reactive power supports proposed by the VVCs for the IICDs better complies with the globally optimal reactive power supports.

Comparing the slope of the VVCs for different objectives, it can be said that the difference of these slopes is not significant at all. For the V2G systems of this study, the slopes of VVCs are the same to two-significant figures. This shows that the capability of an IICD to adhere to a set-point is more linked to the system topology, system impedance, and the location that this inverter has been installed at. Comparing the optimal voltages for these two objectives shows different behaviour under differing objectives. For a LVDS with high penetration of inverter-based RESs, deploying the proposed AVM algorithm results in a significant shift in the system operation when various objectives are taken into account.

The intercepts of the VVCs obtained for loss minimization are usually lower. The reason is that due to the load-to-voltage dependence, the higher voltages lead to the higher active and reactive power demand, higher line currents and higher energy loss. Therefore, with loss minimisation objective, the optimal voltages are lower.

Without updating the VVCs, according to the proposed decentralized resilient AVM algorithm, the voltage control objectives are not satisfied. Sometimes it is possible that without keeping the VVCs up-to-date, the degree of satisfaction of the operator's objective is even lower compared to a system for which the IICDs are not allowed to contribute in the reactive power support provision.

With the practical limitations that should be considered in operation of the inverter-based control devices, the effectiveness of the proposed AVM algorithm is still acceptable according to the results obtained in these studies.

## 6. Acknowledgments

This work was partially supported by the European Commission Ireland, by funding the RESERVE Consortium under Grant 727481 and also supported in part by Science Foundation Ireland (SFI) under the SFI Strategic Partnership Programme Grant Number SFI/15/SPP/ E3125. The opinions, findings and conclusions or recommendations expressed in this material are those of the author(s) and do not necessarily reflect the views of the Science Foundation Ireland